\documentclass[a4paper,pra,twocolumn,showpacs,superscriptaddress,groupedaddress]{revtex4}
\usepackage{ae}
\usepackage{physics}
\usepackage[T1]{fontenc}
\usepackage[ansinew]{inputenc}
\usepackage{amsmath}
\usepackage{amssymb}
\usepackage[caption=false]{subfig}
\usepackage{multirow}
\usepackage{array}
\usepackage[]{graphicx}
\usepackage{wrapfig}
\usepackage{makecell}

\usepackage{color}
\usepackage[colorlinks]{hyperref}
\usepackage{lscape}
\graphicspath{ {./images1/} }

\begin{document}
\title{Monogamy Relations for Multiqubit Systems}
\author{Priyabrata Char}
\email{mathpriyabrata@gmail.com}
\affiliation{Department of Applied Mathematics, University of Calcutta, 92 A.P.C Road, Kolkata 700009, India}
\author{Prabir Kumar Dey}
\email{prabirkumardey1794@gmail.com}
\affiliation{Department of Applied Mathematics, University of Calcutta, 92 A.P.C Road, Kolkata 700009, India}
\author{Amit Kundu}
\email{amit8967@gmail.com}
\affiliation{Department of Applied Mathematics, University of Calcutta, 92 A.P.C Road, Kolkata 700009, India}
\author{Indrani Chattopadhyay}
\email{icappmath@caluniv.ac.in}
\affiliation{Department of Applied Mathematics, University of Calcutta, 92 A.P.C Road, Kolkata 700009, India}
\author{Debasis Sarkar}
\email{dsarkar1x@gmail.com, dsappmath@caluniv.ac.in}
\affiliation{Department of Applied Mathematics, University of Calcutta, 92 A.P.C Road, Kolkata 700009, India}

\begin{abstract}
Recently a new class of monogamy relations (actually, exponentially many) was provided by Christopher Eltschka et al. in terms of squared concurrence. Their approach restricted to the distribution of bipartite entanglement shared between different subsystems of a global state. We have critically analyzed those monogamy relations in three as well as in four qubit pure states using squared negativity. We have been able to prove that in case of pure three qubit states those relations are always true in terms of squared negativity. However, if we consider the pure four qubit states, the results are not always true. Rather, we find opposite behaviour in some particular classes of four qubit pure states where some of the monogamy relations are violated. We have provided analytical and numerical evidences in support of our claim.     
\end{abstract}
\date{\today}
\maketitle
\section{Introduction}
Entanglement is one of the most important ideas in quantum information theory and it is in fact the main form of quantum correlation which shows clear advantages over several aspects of classical theory. Classification and characterization of entanglement have always been a challenging field of research. One important feature of entanglement is that it could be used as a resource that allows one to perform certain quantum information tasks, e.g., dense coding \cite{2}, teleportation \cite{3}, quantum computation \cite{4,5}, etc. Now, as far as the number of parties is concerned, bipartite entanglement is well understood at least for two qubit system, whereas for multipartite systems only few ideas are available. 

Monogamy is one of the most important property of entanglement that provide us the information about the distribution of entanglement in a multipartite system \cite{sev}. Monogamy was possibly first studied by Coffman et al. \cite{6} in terms of squared concurrence. Concurrence is defined as a bipartite measure of entanglement. For a two qubit state $\rho_{AB}$, concurrence is defined by, $C(\rho_{AB})=max\{0,\lambda_1-\lambda_2-\lambda_3-\lambda_4\}$ where $\lambda_1, \lambda_2, \lambda_3, \lambda_4$ are the square root of the eigen values of the matrix $\rho_{AB}((\sigma_y\otimes\sigma_y)\rho^{*}_{AB}(\sigma_y\otimes\sigma_y))$ in decreasing order, $\sigma_y$ is the Pauli spin matrix and $\rho^{*}_{AB}$ is conjugate of $\rho_{AB}$. For pure bipartite states, concurrence can be computed through $C(\rho_{AB})=2\sqrt{det\rho_A}$ where $\rho_A$ is obtained from $\rho_{AB}$ by taking partial trace over the subsystem B. We will use the notation $C_{AB}$ instead of $C(\rho_{AB})$ for any state $\rho_{AB}$. The \textbf{CKW} (Coffman, Kundu, Wootters) inequality \cite{6} is given by,
\begin{equation} 
C^2_{A|BC}\geqslant C^2_{AB}+C^2_{AC}
\end{equation}
 where $C$ denotes the measure of concurrence for a bipartite state. The meaning of the above CKW inequality could be stated as: sum of the amount of entanglement (measured in terms of square of the concurrence) shared between parties A, B and the amount of entanglement shared between the parties A, C can not exceed the amount of entanglement between the parties A and BC. They had also conjectured that the extension of their monogamy relation for n qubit states would be as follows:
\begin{equation}
C^2_{A_1|A_2A_3...A_n}\geqslant C^2_{A_1A_2}+C^2_{A_1A_3}+....+C^2_{A_1A_n}
\end{equation} 
This conjecture later proved by Osborne et al. \cite{7}. Since the introduction of CKW inequality, several works had been done on monogamy where \textbf{CKW} inequality is modified, generalized and also replaced by other entanglement measures \cite{8,9,10,11,12}. All such investigations enables us to understand the entanglement behaviour of composite quantum systems more profoundly. In \cite{13,14}, the authors tried to describe monogamy property without using $\textbf{CKW}$ type inequality \cite{6}. Recently,  C. Eltschka et. al. \cite{1}, provided a new kind of monogamy relation for multipartite (say, $N$ number of parties) $d$ dimensional pure states. They adopt the methodology that any functional relation between measures of entanglement in different subsets of parties could be considered as a monogamy relation because the free distribution of entanglement between different parties has been constrained by it. The monogamy relations in the compact form is \cite{1} given by,
\begin{equation}
\label{original}
\sum_{\Phi\neq S\subset\{1,2,...,N\}}(-1)^{|S\cap T|+1}C^2_{S|S^c}\geqslant0
\end{equation} 
where $\Phi\neq T\subseteq\{1,2,...,N\}$. There are actually $2^N-1$ number of monogamy relations where we find one inequality for each $T$ and when $|T|$ (the cardinality of $T$) is odd we shall get only the trivial inequality $0\geqslant0$. Inspired by their results we have studied in this paper three qubit and four qubit systems through another quantity, the squared negativity.

Negativity is an important measure of entanglement \cite{15}. It is an entanglement monotone and invariant under local unitary operations. The negativity is a rare bipartite entanglement measure which is easy to compute for pure as well for mixed bipartite states. From Peres criterion \cite{16}, it is known that for a separable state partial transpose of its density matrix will also be a density matrix. Partial transpose in general preserve hermiticity but not positivity. Thus after taking partial transpose on a density matrix representing a bipartite state, if we obtain at least one negative eigen value, then we could certainly say that the state is an entangled state. The definition of negativity for a bipartite state $\rho_{AB}$ (pure or mixed) is given by,
\begin{equation}
N(\rho_{AB})=\frac{\|\rho^{t_A}_{AB}\|_1-1}{2}
\end{equation}  
where $\|X\|_1=tr\sqrt{XX^\dagger}$ and partial transposition is taken with respect to subsystem A. In other words, the negativity is the absolute sum of negative eigenvalues of $\rho_{AB}^{t_A}$ and it measures how much $\rho_{AB}^{t_A}$ fails to be a positive definite matrix. We will use the notation $N_{AB}$ instead of $N(\rho_{AB})$.\\

We have organized our paper as follows: In section II, we will discuss motivation of our work. In section III and IV we will discuss monogamy relations for three qubit and four qubit pure states respectively. Section V ended with conclusion.

\section{General Motivition}
The generalized $T$ inversion map \cite{1} is, 
\begin{equation}
\label{old}
\mathcal{I}_{T}(\rho)=\sum_{S\subseteq\{1,2,...,N\}}(-)^{|S\cap T|}(Tr_{S^c}\rho)\otimes I_{S^c}
\end{equation}
where $T$ is any subset of $\{1,2,...,N\}$. Using positivity property of $\mathcal{I}_{T}(.)$, for two semi definite positive operator $M_1$ and $M_2$ one has 
\begin{equation}
\label{old1}
Tr_S[M_1\mathcal{I}_{T}(M_2)]\geq 0
\end{equation}
As $Tr_S[(M_1)Tr_{S^c}(M_2)]=Tr_S[Tr_{S^c}(M_1)Tr_{S^c}(M_2)]$  putting equation \eqref{old} in \eqref{old1} one will get 
\begin{equation}
\label{shadow}
\sum_{S\subseteq\{1,2,...,N\}}(-)^{|S\cap T|}Tr_{S}[Tr_{S^c}(M_1)Tr_{S^c}(M_2)]\geq 0 
\end{equation}
where $T$ is any subset of $\{1,2,...,N\}$. This inequality is called shadow inequality \cite{motivation 1,motivation 2}.

Now, if one consider $M_1=M_2=\ket{\psi_{N,D}}$ an $N$ partite $D$ dimensional pure state then one can directly get the monogamy inequalities,
\begin{equation}
\label{monogamy}
\sum_{\Phi\neq S\subseteq\{1,2,...,N\}}(-1)^{|S\cap T|+1}C^2_{S|S^c}\geqslant0
\end{equation} 
where $\Phi\neq T\subseteq\{1,2,...,N\}$ and here $C_{S|S^c}$ is concurrence of the pure state along the bipartition. So, the relations \eqref{monogamy} are direct consequences of shadow inequality or rather the algebraic property of generalized $T$ inverter.

Again, the shadow enumerator polynomial \cite{motivation 1} is,
 \begin{equation}
S_{M_1M_2}(x,y)=\sum_{j=0}^{N} S_j(M_1M_2) x^{N-j} y^j
\end{equation} where the coefficient is defined as follows
\begin{equation}
S_j(M_1M_2)=\sum_{|T|=j}\sum_{S\subseteq\{1,...,N\}}(-1)^{|S\cap T^c|}\mathcal{A}_{S}^{/}(M_1,M_2)
\end{equation} (the first sum is over all subset of size $j$) and
 $\mathcal{A}_{S}^{/}(M_1,M_2)=Tr_S[Tr_{S^c}(M_1)Tr_{S^c}(M_2)]$. 
If in particular $M_1=M_2=\ket{\psi_{N,D}}$ then the inequalities \eqref{monogamy} will imply that $S_j(\ket{\psi_{N,D}})\geq 0$.

Further, $S_j(M_1M_2)$ can be written in terms of coefficient of Shor-Laflamme enumerator \cite{motivation 3} which is 
\begin{equation}
S_j(M_1M_2)=\sum_{l=0}^{N} K_{N-j}(l;N) A_{l}^{/}(M_1,M_2) 
\end{equation} 
where $K_{N-j}(l;N)$ is the Krawtchouk polynomial $$K_m(l;N)=\sum_{\alpha}(-1)^\alpha \binom{n-l}{m-\alpha}\binom{l}{\alpha}$$
Now, when $M_1=M_2=\ket{\psi_{N,D}}$, then $A_{l}^{/}(M_1,M_2)=\binom{N}{l} D^{-min(l,N-l)}.$

Therefore, 
\begin{equation}
S_j(\ket{\phi_{N,D}})=\sum_{l=0}^{N} K_{N-j}(l;N)\binom{N}{l} D^{-min(l,N-l)} 
\end{equation} 
If for a pure state $\ket{\psi_{N,D}}$, $S_j(\ket{\psi_{N,D}})$ becomes negative then an Absolute Maximally Entangled (AME) \cite{motivation 3} state on $N$ parties having $D$ dimension can not exist as it will contradict $S_j(\ket{\psi_{N,D}})\geq 0$. 

A particular example is $\ket{\psi_{4,2}}$, where $S_0(\ket{\psi_{4,2}})=\sum_{l=0}^{4} (-1)^k\binom{4}{l} 2^{-min(l,4-l)}=-\frac{1}{2}<0$. Therefore, there does not exist a $4$ partite $2$ local dimensional AME state \cite{motivation 3}.

The inequalities \eqref{monogamy} are very important class of monogamy inequalities, as because in one hand, it is derived from an algebraic property of generalized $T$ inverter and on the other hand, it helps one in excluding the existence of AME states in $N$ partite $D$ local dimensions. A simple question that arises from their work is whether this type of monogamy holds for other entanglement measures or not. In our work, we have examined the above set of monogamy relations using negativity as an entanglement measure for three and four qubit pure states. 

\section{Monogamy relations for three qubit pure states}
We start this section with a relation between negativity and concurrence. 

\textbf{Theorem 1.} \cite{10} For an $N$ partite pure state $\ket{\psi_{A_1A_2...A_N}}$ in a $2\otimes2\otimes...\otimes2$($N$ times) system the negativity of bipartition $A_1|A_2...A_N$ is half of its concurrence, i.e., $N_{A_1|A_2...A_N}=\frac{1}{2}C_{A_1|A_2...A_N}$. \\
Proof is given in appendix 3.

We will use the above theorem to form monogamy relations for three and four qubit systems from relations \eqref{original} with respect to squared negativity.
For a three qubit pure state, from monogamy relations \eqref{original}, we have, 
\begin{equation}
\label{5}
\sum_{\phi\neq S\subset\{1,2,3\}}(-1)^{|S\cap T|+1}C^2_{S|S^c}\geq 0
\end{equation}
where we will get one inequality for each $\Phi\neq T\subseteq\{1,2,3\}$, i.e., total $2^3-1=7$ monogamy relations. When $|T|$ is odd we shall obtain trivial inequality $0\geq0$. Expanding \eqref{5} for $T=\{1,2\},~~T=\{1,3\},~~T=\{2,3\}$ we get, respectively
\begin{equation}
\label{6}
C^2_{1|23}+C^2_{2|13}\geq C^2_{3|12}
\end{equation}
\begin{equation}
\label{7}
C^2_{1|23}+C^2_{3|12}\geq C^2_{2|13}
\end{equation}
\begin{equation}
\label{8}
C^2_{2|13}+C^2_{3|12}\geq C^2_{1|23}
\end{equation}
Now, using theorem 1 for $2\otimes 2\otimes 2$ dimensional pure states, we have, $C_{i|jk}=2\times N_{i|jk}$ and thus from \eqref{6},\eqref{7},\eqref{8} we can write,
\begin{equation}
\label{9}
N^2_{1|23}+N^2_{2|13}\geq N^2_{3|12}
\end{equation}
\begin{equation}
\label{10}
N^2_{1|23}+N^2_{3|12}\geq N^2_{2|13}
\end{equation}
\begin{equation}
\label{11}
N^2_{2|13}+N^2_{3|12}\geq N^2_{1|23}
\end{equation}
The above three monogamy inequalities can also be written compactly as 
\begin{equation}
\sum_{\phi\neq S\subset\{1,2,3\}}(-1)^{|S\cap T|+1}N^2_{S|S^c}\geq 0
\end{equation}
where one inequality is associated for each $\Phi\neq T\subset\{1,2,3\}$, i.e., total $(2^3-2)=6$ inequalities. When $|T|$ is odd we shall get only the trivial inequality $0\geq0$ \cite{1}. Thus, theorem 1 completely determines the monogamy relations in terms of squared negativity from the relation \eqref{5}. Next, we will consider pure four qubit states and observe whether it is similar to that of three qubit case or not. 
\section{Monogamy relations for four qubit pure states} 
For a four qubit pure state relations \eqref{original} looks like
\begin{equation}
\label{13}
\sum_{\phi\neq S\subset\{1,2,3,4\}}(-1)^{|S\cap T|+1}C^2_{S|S^c}\geq 0
\end{equation}
where one inequality is associated for each $\Phi\neq T\subseteq\{1,2,3,4\}$, i.e., total $2^4-1=15$ monogamy relations, out of which eight are trivial inequalities $0\geq0$ when $|T|$ is an odd number. The inequalities \eqref{13} are given in details in appendix 1. We now state another relation between concurrence and negativity in the following theorem.

\textbf{Theorem 2.}
For an $N$ partite pure state $\ket{\psi_{A_1A_2...A_N}}$ in a $d_1\otimes d_2\otimes ...\otimes d_n$ dimensional system where each $d_i>2$ $\forall i=1,2,..., n$, $N_{A_1|A_2...A_N}>\frac{1}{2}C_{A_1|A_2...A_N}$ .\\
Proof is given in appendix 3. 

As stated in theorem 2 the replacement of concurrence by negativity in the relations \eqref{13} is not always possible like in the three qubit case, since in some expressions, the focus party is of dimension $4$, hence theorem 1 will not be applicable to such cases.

We now denote $\delta_i$, $\forall i=1,2,...,15$ as follows,
\begin{equation}
\label{14}
\delta_i=\sum_{\phi\neq S\subset\{1,2,3,4\}}(-1)^{|S\cap T|+1}N^2_{S|S^c}
\end{equation}
where we obtain, for each $\Phi\neq T\subseteq\{1,2,3,4\}$, total $2^4-1=15$ expression. When $|T|$ is odd we shall get zero in the right hand side of \eqref{14}. We take the non zero expressions as $\delta_1,\delta_2,...\delta_7$ and $\delta_8=\delta_9=...=\delta_{15}=0$. Expansion of expressions \eqref{14} are given in appendix 1. 

Whenever $\delta_{i}\geq0$, $\forall i=1,2,...,7$ we have the relations \eqref{25}-\eqref{31}, given in appendix 1, are true. As there exists infinitely many SLOCC inequivalent classes for four qubit pure states, we will consider the four qubit generic class \cite{17} and other important four qubit classes to check the sign of $\delta_i$'s $\forall i=1,2,...,7$.

\subsection{MONOGAMY RELATIONS IN SOME PARTICULAR CLASSES OF FOUR QUBIT PURE STATES}
\textsl{Generic Class:} The generic class of pure states is dense under SLOCC in four qubit state space. It even contains uncountable SLOCC inequivalent subclasses \cite{18}. We denote this class by $\mathcal{A} $ and is defined as \\
\begin{equation*}
\begin{split}
\mathcal{A}=\{au_1+bu_2+cu_3+du_4 \quad|\quad a,b,c,d\in \mathbb{C}\\ \quad \text{and} \quad|a|^2+|b|^2+|c|^2+|d|^2=1 \}
\end{split}
\end{equation*}
where $u_1\equiv|\Phi^+\rangle|\Phi^+\rangle$, $u_2\equiv|\Phi^-\rangle|\Phi^-\rangle$, $u_3\equiv|\Psi^+\rangle|\Psi^+\rangle$, $u_4\equiv|\Psi^-\rangle|\Psi^-\rangle$, $|\Phi^{\pm}\rangle=\frac{|00\rangle\pm|11\rangle}{\sqrt{2}}$ and  $|\Psi^{\pm}\rangle=\frac{|01\rangle\pm|10\rangle}{\sqrt{2}}$
We now consider two special subclasses of generic class \cite{18} of four qubit pure states  
\begin{equation*}
\begin{split}
\mathcal{B}=\{au_1+au_2+cu_3+cu_4 \quad|\quad a,c\in \mathbb{C}\\ \quad \text{and} \quad2(|a|^2+|c|^2)=1 \}
\end{split}
\end{equation*}
and
\begin{equation*}
\begin{split}
\mathcal{D}=\{au_1+bu_2+cu_3+du_4 \quad|\quad a,b,c,d\in \mathbb{R}\\ \quad \text{and} \quad|a|^2+|b|^2+|c|^2+|d|^2=1 \}
\end{split}
\end{equation*}
For states in subclass $\mathcal{B}$ we have $N_{1|234}=N_{2|134}=N_{3|124}=N_{4|123}=\frac{1}{2},\quad N_{12|34}=N_{14|23}=|a|^2+|c|^2+4|ac|$ $\text{and}$ $N_{13|24}=|a^2-c^2|. \quad$  \\
So, $\delta_1=\delta_2=\delta_5=\delta_6=|a^2-c^2|^2\geq0 $,\\
$\delta_3=\delta_4=|a|^4+|c|^4+16|ac|[|a|^2+|c|^2]+2[18|ac|^2+Re(a^2{c^*}^2)]\geq 0$,  as  $Re(a^2{c^*}^2)\leq|a^2{c^*}^2|=|a^2c^2|$.\\
Due to the difficulties in finding the sign of $\delta_7$, numerical simulation (FIG. 1) have been performed with $10^5$ random pure states from class $\mathcal{B}$, which clearly shows that $\delta_7<0$ for most of the cases.
\begin{figure}
\centering
\includegraphics[scale=.6]{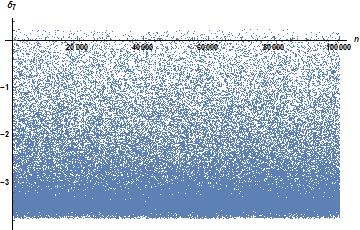}
\caption{$\delta_7$ for states in $\mathcal{B}$}
\end{figure}

In particular, if we take $a$ and $c$ as real numbers then we have obtained the graph of $\delta_7 $  vs  $ a$ (FIG. 2).\\
\begin{figure}
\centering
\includegraphics[scale=.6]{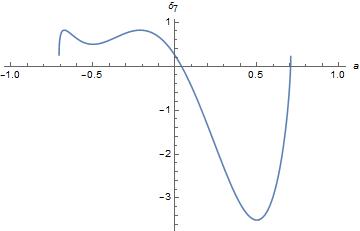}
\caption{$a$ vs $\delta_7$ for state in $\mathcal{B}$}
\end{figure}
For the states in subclass  $\mathcal{D}$ (see details in appendix 2) due to the difficulty in computation of sign of $\delta_i,$ $\forall i=1,2,...,7$, we present numerical evidences using $10^5$ random pure states from class $\mathcal{D}$ which shows  $\delta_1=\delta_2\geq0$ (FIG. 3), 
\begin{figure}
\centering
\includegraphics[scale=.6]{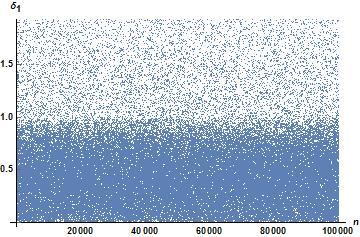}
\caption{$\delta_1$ for states in $\mathcal{D}$}
\end{figure}
 Also, $\delta_3=\delta_4\geq0 $ $\&$ $\delta_5=\delta_6\geq0  $ (FIG. 8 \& FIG. 9 in appendix 2) in all cases. But, numerical evidences for $\delta_7$ (FIG. 4) shows that it is negative for most of the cases except for a small number.
 % So, we observe that the monogamy relations \eqref{25}-\eqref{30} hold but \eqref{31} does not hold for the subclass $\mathcal{B}$ and $\mathcal{D}$ of four qubit generic class.
 \\
\begin{figure}
\centering
\includegraphics[scale=.6]{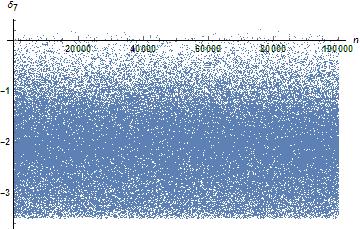} 
\caption{$\delta_7$ for states in $\mathcal{D}$}
\end{figure}
\begin{figure}
\centering
\includegraphics[scale=.6]{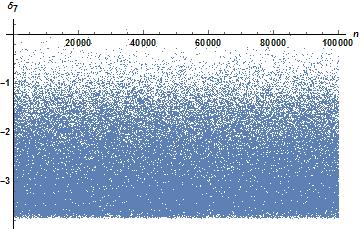} 
\caption{$\delta_7$ for cluster states}
\end{figure}

\textsl{Cluster States:} Cluster states are used in quantum nonlocality test \cite{19}, quantum error correction code \cite{20}, etc. Four qubit cluster states \cite{21} can be written as \\$\ket{\psi}=a\ket{0000}+b\ket{0011}+c\ket{1100}-d\ket{1111} $\\ where $ a,b,c,d\in \mathcal{C}$ and $|a|^2+|b|^2+|c|^2+|d|^2=1$. Calculating negativity for this state we observe that $\delta_i\geq0$ $ \forall i=1,2,...,6$ (see appendix 2). For $\delta_7$ numerical simulation with $10^5$ random states from this class have been performed [FIG. 5],  which shows that for most of the cases $\delta_7<0$.% So, \eqref{25}-\eqref{30} are always true for cluster state but \eqref{31} is not true again.
\\

\textsl{Dicke States:} A four qubit Dicke state \cite{22} is given by, $$\ket{S(4,k)}=\sqrt{\frac{k!(4-k)!}{4!}}\sum_{permutation}\ket{0}^{\otimes(4-k)}\ket{1}^{\otimes k}$$ where the summation is over all possible permutations of the product state having $k(\leq 4)$ qubit in excited state $\ket{1}$ and remaining $(4-k)$ qubits are in ground state. $\ket{S(4,0)}=\ket{0000}$ and $\ket{S(4,4)}=\ket{1111}$, are separable states. $\ket{S(4,1)}=\ket{W}$ and $\ket{S(4,3)}=\ket{\tilde{W}}$. For $\ket{W}$ and $\ket{\tilde{W}}$ we get, $\delta_i=\frac{1}{4}$ $ $ $\forall i=1,2,...,6 $ and $\delta_7=0$ (See Appendix 2).
% So \eqref{25}-\eqref{31} are true for $\ket{W}$ and $\ket{\tilde{W}}$. 
 When $k=2$, we get $\ket{S(4,2)}=(\ket{0011}+\ket{1100}+\ket{0110}+\ket{1001}+\ket{1010}+\ket{0101})/\sqrt{6}$. 
 For this state, we have $\delta_i=\frac{25}{36}$ $>0$, $\forall i=1,2,...,6 $ and this time, $\delta_7=-\frac{13}{12}$ $<0$ (See Appendix 2). %Hence for $\ket{S(4,2)}$ \eqref{25}-\eqref{30} are true whereas \eqref{31} is violated in this case.
 \\
 
\textsl{Generalized GHZ State:}
Four qubit generalized GHZ state is
$\ket{GGHZ}=a\ket{0000}+b\ket{1111}$ where, $a,b\in\mathbb{C}$ and $|a|^2+|b|^2=1$. Simple calculations have yielded that $N_{1|234}=N_{2|134}=N_{3|124}=N_{4|123}=N_{12|34}=N_{13|24}=N_{14|23}=|ab|$. Hence $\delta_i=|ab|^2>0,$ $\forall i=1,2,...,7$. %For Generalized GHZ state monogamy relations \eqref{25}-\eqref{31} are all satisfied. 
\\

\textsl{Generalized W State:}
Four qubit generalized W state is given by, $\ket{GW}=a\ket{0001}+b\ket{0010}+c\ket{0100}+d\ket{1000}$ where $a,b,c,d\in\mathbb{C}$ and $|a|^2+|b|^2+|c|^2+|d|^2=1$. Simple calculations (see appendix 2) have revealed that $\delta_i\geqslant0$ $\forall i=1,2,...,6$ and $\delta_7=0$. % So \eqref{25}-\eqref{31} are always true for generalized W state also.
 Obviously the results for $W$ state can be derived directly from the generalized $W$ state. 
\subsection{Monogamy Relations In Superposition Of Some Pure States }
\textsl{Superposition of $\ket{W}$ and $\ket{\tilde{W}}$ states:}
Consider the superposition of $\ket{W} \& \ket{\tilde{W}}$ as $\ket{\psi}=a\ket{\tilde{W}}+be^{i\theta}\ket{W}$ where $a,b\in(0,1),$ $a^2+b^2=1\&$ $\theta\in[0,2\pi].$
Here, $N_{1|234}=N_{2|134}=N_{3|124}=N_{4|123}=\frac{1}{4}\sqrt{3+4a^2b^2}$ and $N_{12|34}=N_{13|24}=N_{14|23}=\frac{1}{2}$. Therefore, we have, $\delta_i=\frac{1}{4}>0$ $\forall i=1,2,...,6$ and $\delta_7=a^2b^2>0$. % So \eqref{25}-\eqref{31} are satisfied for this case.\\
\\

\textsl{Superposition of $\ket{GW}$ and $\ket{0000}$:} Suppose, $\ket{\psi}=\sqrt{p}\ket{GW}+\sqrt{1-p}\ket{0000}$ where $0<p<1$, $\ket{GW}=a\ket{0001}+b\ket{0010}+c\ket{0100}+d\ket{1000}$, $a,b,c,d\in\mathbb{C}$ such that  $|a|^2+|b|^2+|c|^2+|d|^2=1. $
For this case we have, $\delta_i\geq 0, ~\forall i=1,2,...,6$ and $\delta_7=0 $  (see appendix 2). %Hence \eqref{25}-\eqref{30} are satisfied and equality occurs for relation \eqref{31}.
\\

\textsl{Superposition of $\ket{GGHZ}$ and $\ket{W}$:} 
Consider, $\ket{\psi}=c_1(a_1\ket{0000}+b_1\ket{1111})+c_2(\ket{0001}+\ket{0010}+\ket{0100}+\ket{1000})/2$, $a_1,b_1,c_1,c_2\in\mathbb{C}$ such that $|a_1|^2+|b_1|^2=1$ and $|c_1|^2+|c_2|^2=1$. Considering $c_1a_1, c_1b_1, c_2$ as $a,b,c$ respectively $\ket{\psi}=a\ket{0000}+b\ket{1111}+\frac{c}{2}(\ket{0001}+\ket{0010}+\ket{0100}+\ket{1000})$ where $a,b,c\in\mathbb{C}$ such that $|a|^2+|b|^2+|c|^2=1$. For this case we have, $\delta_i\geq 0,$ $\forall i=1,2,...,6$ (see appendix 2). Due to the difficulty in computation of sign of $\delta_7$,  numerical evidence [FIG. 6] is presented using $10^5$ random pure states from this class. The figure 6 clearly explains that $\delta_7$ can be positive, negative or even zero for  states in this superposed class [FIG. 6].\\
\begin{figure}
\centering
\includegraphics[scale=.6]{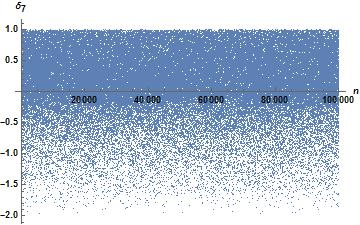}
\caption{$\delta_7$ for superposition states of $\ket{GGHZ}$ and $\ket{W}$}
\end{figure}
\begin{figure}
\centering
\includegraphics[scale=.6]{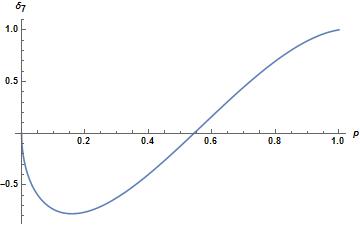}
\caption{$p$ vs $\delta_7$ for superposition of $\ket{GHZ}$ and $\ket{W}$}
\end{figure}
Particularly assuming, $a=b=\sqrt{p/2} $  and $c=\sqrt{1-p}$ where $p\in(0,1)$ and we have obtained $p$ vs $\delta_7$ graph [FIG. 7].% Hence, in any case \eqref{25}-\eqref{30} true, but relation \eqref{31} is not always true.
\\

For the four qubit case, we consider different physically important pure states and some subclasses of generic class. It is observed that the relations \eqref{25}-\eqref{30} are well satisfied for all the mentioned classes and states in this paper, but peculiar behaviour of the relation \eqref{31} have been noticed here. We have proved that the relation \eqref{31} holds for generalized GHZ state, generalized W state, superposition of $\ket{W}$ and $\ket{\tilde{W}}$ state, superposition of generalized W and ground state $\ket{0000}$, whereas violation is observed in subclasses $\mathcal{B},\mathcal{D}$ of four qubit pure generic class, Dicke $\ket{S(4,2)}$ and by Cluster state. The most counter-intuitive result has been noticed through the superposition of W state and generalized GHZ state where we see \eqref{31} has been violated as well as satisfied for large number of random states. Another important observation of our work enlighten the fact that superposition of states also plays a crucial role on status of \eqref{31}, contrary to \eqref{25}-\eqref{30}. $\delta_7=0$ for $\ket{W}$ and $\ket{\tilde{W}}$ but for their superposition $\delta_7> 0$. Similar, peculiar behaviour of \eqref{31} has been observed for superposition of $\ket{GGHZ}$ and $\ket{W}$, where $\delta_7$ changes sign near $p=0.55$ (approx) [FIG. 7], i.e., in this case, \eqref{31} violated and satisfied depending on the value of $p$. 
\section{Conclusion}
In conclusion, we have analyzed a new set of monogamy relations in terms of squared negativity for three qubit and four qubit pure states. With the help of theorem 1 we have proved three monogamy relations \eqref{9}-\eqref{11} analytically and compactly. We can write them as $$\sum_{\phi\neq S\subset\{1,2,3\}}(-1)^{|S\cap T|+1}N^2_{S|S^c}\geq 0$$ where we will get one inequality for each $\Phi\neq T\subset\{1,2,3\}$. In four qubit case  for squared negativity, we see that the six relations \eqref{25}-\eqref{30} plus eight trivial inequalities ($0\geq 0$), i.e., total fourteen monogamy relations of type $$\sum_{\phi\neq S\subset\{1,2,3,4\}}(-1)^{|S\cap T|+1}N^2_{S|S^c}\geq 0$$ where we will get one inequality for each $\Phi\neq T\subset\{1,2,3,4\}$ are always true in all the considered cases of this paper. We have observed that for three qubit case when $T=\{1,2,3\}$, we get a trivial inequality $0\geq 0$ and in four qubit case when $T=\{1,2,3,4\}$, the corresponding inequities \eqref{31} show different behaviour for different classes. That is why we have excluded the case when $T$ is the set of all parties. We conjecture that for $N$ qubit pure states the monogamy relations are $$\sum_{\phi\neq S\subset\{1,2,...,N\}}(-1)^{|S\cap T|+1}N^2_{S|S^c}\geq 0$$ where we will get one inequality for each $\Phi\neq T\subset\{1,2,...,N\}$, i.e., total $(2^N-2)$ inequalities and when $|T|$ is odd we will get the trivial inequality $0\geq0$. We hope our result will provide further insight on entanglement distribution of multipartite systems and could be applied on possible areas of quantum key distributions and quantum cryptography.
\section*{ACKNOWLEDGEMENTS}
Priyabrata Char acknowledges the support from Department of Science \& Technology(Inspire), New Delhi, India, Prabir Kumar Dey acknowledges the support from UGC, New Delhi, and Amit Kundu acknowledges the support from CSIR, New Delhi, India. The authors D. Sarkar and I. Chattopadhyay acknowledges it as QuEST initiatives.

\vspace{10in}
\section*{Appendix 1}
\begin{widetext}
\begin{align}
\label{18}
C^2_{1|234}+C^2_{2|134}+C^2_{13|24}+C^2_{14|23} &\geq C^2_{3|124}+C^2_{4|123}+C^2_{12|34}\hspace{0.2in}for\hspace{5pt} T=\{1,2\}\\
\label{19}
C^2_{3|124}+C^2_{4|123}+C^2_{13|24}+C^2_{14|23} &\geq C^2_{1|234}+C^2_{2|134}+C^2_{12|34}\hspace{0.2in}for\hspace{5pt} T=\{3,4\}\\
\label{20}
C^2_{1|234}+C^2_{3|124}+C^2_{12|34}+C^2_{14|23} &\geq C^2_{4|123}+C^2_{2|134}+C^2_{13|24}\hspace{0.2in}for\hspace{5pt} T=\{1,3\}\\
\label{21}
C^2_{4|123}+C^2_{2|134}+C^2_{12|34}+C^2_{14|23} &\geq C^2_{1|234}+C^2_{3|124}+C^2_{13|24}\hspace{0.2in}for\hspace{5pt} T=\{2,4\}\\
\label{22}
C^2_{1|234}+C^2_{4|123}+C^2_{12|34}+C^2_{13|24} &\geq C^2_{2|134}+C^2_{3|124}+C^2_{14|23}\hspace{0.2in}for\hspace{5pt} T=\{1,4\}\\
\label{23}
C^2_{2|134}+C^2_{3|124}+C^2_{12|34}+C^2_{13|24} &\geq C^2_{1|234}+C^2_{4|123}+C^2_{14|23}\hspace{0.2in}for\hspace{5pt} T=\{2,3\}\\
\label{24}
C^2_{1|234}+C^2_{2|134}+C^2_{3|124}+C^2_{4|123} &\geq C^2_{12|34}+C^2_{13|24}+C^2_{14|23}\hspace{0.2in}for\hspace{5pt} T=\{1,2,3,4\}
\end{align}
%\hspace{1.05in} $0\geq 0$ when $|T|$ is odd number
%\end{widetext}
%\begin{widetext}
\begin{align*}
\delta_1 &= N^2_{1|234}+N^2_{2|134}+N^2_{13|24}+N^2_{14|23}-N^2_{3|124}-N^2_{4|123}-N^2_{12|34}\hspace{0.2in}for\hspace{5pt} T=\{1,2\}\\
\delta_2 &= N^2_{3|124}+N^2_{4|123}+N^2_{13|24}+N^2_{14|23}-N^2_{1|234}-N^2_{2|134}-N^2_{12|34}\hspace{0.2in}for\hspace{5pt} T=\{3,4\}\\
\delta_3 &= N^2_{1|234}+N^2_{3|124}+N^2_{12|34}+N^2_{14|23}-N^2_{4|123}-N^2_{2|134}-N^2_{13|24}\hspace{0.2in}for\hspace{5pt} T=\{1,3\}\\
\delta_4 &= N^2_{4|123}+N^2_{2|134}+N^2_{12|34}+N^2_{14|23}-N^2_{1|234}-N^2_{3|124}-N^2_{13|24}\hspace{0.2in}for\hspace{5pt} T=\{2,4\}\\
\delta_5 &= N^2_{1|234}+N^2_{4|123}+N^2_{12|34}+N^2_{13|24}-N^2_{2|134}-N^2_{3|124}-N^2_{14|23}\hspace{0.2in}for\hspace{5pt} T=\{1,4\}\\
\delta_6 &= N^2_{2|134}+N^2_{3|124}+N^2_{12|34}+N^2_{13|24}-N^2_{1|234}-N^2_{4|123}-N^2_{14|23}\hspace{0.2in}for\hspace{5pt} T=\{2,3\}\\
\delta_7 &= N^2_{1|234}+N^2_{2|134}+N^2_{3|124}+N^2_{4|123}-N^2_{12|34}-N^2_{13|24}-N^2_{14|23}\hspace{0.2in}for\hspace{5pt} T=\{1,2,3,4\}\\
\delta_8&=\delta_9=...=\delta_{15}=0 \hspace{0.1in} when \hspace{0.05in} |T| \hspace{0.05in}is\hspace{0.05in} odd\hspace{0.05in} number. 
\end{align*}
%\end{widetext}
%\begin{widetext}
\begin{align}
\label{25}
N^2_{1|234}+N^2_{2|134}+N^2_{13|24}+N^2_{14|23} &\geq N^2_{3|124}+N^2_{4|123}+N^2_{12|34}\hspace{0.2in}for\hspace{5pt} T=\{1,2\}\\
\label{26}
N^2_{3|124}+N^2_{4|123}+N^2_{13|24}+N^2_{14|23} &\geq N^2_{1|234}+N^2_{2|134}+N^2_{12|34}\hspace{0.2in}for\hspace{5pt} T=\{3,4\}\\
\label{27}
N^2_{1|234}+N^2_{3|124}+N^2_{12|34}+N^2_{14|23} &\geq N^2_{4|123}+N^2_{2|134}+N^2_{13|24}\hspace{0.2in}for\hspace{5pt} T=\{1,3\}\\
\label{28}
N^2_{4|123}+N^2_{2|134}+N^2_{12|34}+N^2_{14|23} &\geq N^2_{1|234}+N^2_{3|124}+N^2_{13|24}\hspace{0.2in}for\hspace{5pt} T=\{2,4\}\\
\label{29}
N^2_{1|234}+N^2_{4|123}+N^2_{12|34}+N^2_{13|24} &\geq N^2_{2|134}+N^2_{3|124}+N^2_{14|23}\hspace{0.2in}for\hspace{5pt} T=\{1,4\}\\
\label{30}
N^2_{2|134}+N^2_{3|124}+N^2_{12|34}+N^2_{13|24} &\geq N^2_{1|234}+N^2_{4|123}+N^2_{14|23}\hspace{0.2in}for\hspace{5pt} T=\{2,3\}\\
\label{31}
N^2_{1|234}+N^2_{2|134}+N^2_{3|124}+N^2_{4|123} &\geq N^2_{12|34}+N^2_{13|24}+N^2_{14|23}\hspace{0.2in}for\hspace{5pt} T=\{1,2,3,4\}
\end{align}
\end{widetext}
\section*{Appendix 2}
The subclass of four qubit pure generic state $\mathcal{D}$ is
$\mathcal{D}=\{au_1+bu_2+cu_3+du_4 \quad|\quad a,b,c,d\in \mathbb{R}\\ \quad \text{and} \quad|a|^2+|b|^2+|c|^2+|d|^2=1 \}\\ $
For the states in subclass  $\mathcal{D}$ we have,\\
$N_{1|234}=N_{2|134}=N_{3|124}=N_{4|123}=\frac{1}{2}$ ,\\
$ N_{13|24}=\{|(a+b)^2-(c+d)^2|+|(a-b)^2-(c-d)^2|+|(a+c)^2-(b+d)^2|+|(a-c)^2+(b-d)^2|+|(a+d)^2-(b+c)^2|+|(a-d)^2-(b-c)^2|\}/4$ ,\\ 
$N_{14|23}=\{|(a+b)^2-(c-d)^2|+|(a-b)^2-(c+d)^2|+|(a+c)^2-(b-d)^2|+|(a-c)^2+(b+d)^2|+|(a+d)^2-(b-c)^2|+|(a-d)^2-(b+c)^2|\}/4$ , \\ 
 $N_{12|34}=|ab|+|ac|+|ad|+|bc|+|bd|+|cd|$ .\\
$\delta_1=\delta_2=N_{13|24}^2+N_{14|23}^2-N_{12|34}^2 \hspace{.1cm}, \\ 
\delta_3=\delta_4=N_{12|34}^2+N_{14|23}^2-N_{13|24}^2 \hspace{.1cm}, \\
\delta_5=\delta_6=N_{12|34}^2+N_{13|24}^2-N_{14|23}^2 \hspace{.1cm},\\
\delta_7=1-N_{12|34}^2-N_{13|24}^2-N_{14|23}^2 \hspace{.1cm}. $ $ $\\
The numerical simulations using $10^5$ pure random states form class $\mathcal{D}$  
shows that $\delta_3=\delta_4\geq0$ (FIG. 8) and $\delta_5=\delta_6\geq0$ (FIG. 9).
\begin{figure}
\centering
\includegraphics[scale=.6]{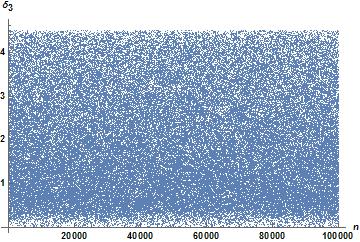}
\caption{$\delta_3$ for state in subclass $\mathcal{D}$}
\end{figure}
\begin{figure}
\centering
\includegraphics[scale=.6]{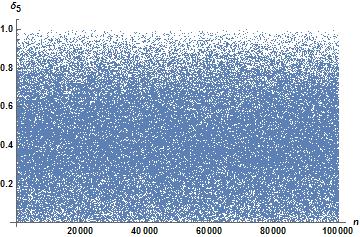}
\caption{$\delta_5$ for state in subclass $\mathcal{D}$}
\end{figure} \\

Four qubit cluster state is
$\ket{\psi}=a\ket{0000}+b\ket{0011}+c\ket{1100}-d\ket{1111} $ where $ a,b,c,d\in \mathcal{C}$ and $|a|^2+|b|^2+|c|^2+|d|^2=1$.
Negativities of cluster state are $N_{12|34}=|bc+ad| $ ,\\ 
$N_{13|24}=N_{14|23}=|ab|+|ac|+|ad|+|bc|+|bd|+|cd|$ ,\\
$N_{1|234}=N_{2|134}=\sqrt{(|a|^2+|b|^2)(|c|^2+|d|^2)}$ ,\\
$N_{3|124}=N_{4|123}=\sqrt{(|a|^2+|c|^2)(|b|^2+|d|^2)}$ .\\
$\delta_3=\delta_4=N_{12|34}^2+N_{14|23}^2-N_{13|24}^2=|bc+ad|^2\geq0 \hspace{.1cm},\\
\delta_5=\delta_6=N_{12|34}^2+N_{13|23}^2-N_{14|24}^2=|bc+ad|^2\geq0 \hspace{.1cm},\\ $
$\delta_1=4(|ac|^2+|bd|^2)+(|bc|^2+|ad|^2)+2(|bcad|-Re(bca^*d^*))+2L\geq0 $ , \\
$\delta_2=4(|ab|^2+|cd|^2)+(|bc|^2+|ad|^2)+2(|bcad|-Re(bca^*d^*))+2L\geq0 $ [$\because$ $|bc||ad|\geq Re(bca^*d^*) $] ,\\
 where $L$ is sum of product of $\{|ab|,|ac|,|ad|,|bc|,|bd|,|cd|\}$ taken two at a time except the product $|bc||ad|$. \\
\\
The $\ket{W}$ and $\ket{\tilde{W}}$ states are \\
$\ket{W}=\frac{1}{2}(\ket{0001}+\ket{0010}+\ket{0100}+\ket{1000})$\\
$\ket{\tilde{W}}=\frac{1}{2}(\ket{1110}+\ket{1101}+\ket{1011}+\ket{0111})$\\
Negativities of $\ket{W}$ and $\tilde{W}$ states are $N_{1|234}=N_{2|134}=N_{3|124}=N_{4|123}=\frac{\sqrt{3}}{4}$ and $N_{12|34}=N_{13|24}=N_{14|23}=\frac{1}{2}$. Hence, $\delta_i=\frac{1}{4}>0$ $ \forall i=1,2,...,6$, but $\delta_7=0$. The negativities of $\ket{S(4,2)}$ among different bipartition are $N_{1|234}=N_{2|134}=N_{3|124}=N_{4|123}=\frac{1}{2}$ and $N_{12|34}=N_{13|24}=N_{14|23}=\frac{5}{6}$. Thus, $\delta_i=\frac{25}{36}$ $>0$, $\forall i=1,2,...,6 $ and $\delta_7=-\frac{13}{12}$ $<0$.\\
\\
Generalized W state is \\
$\ket{GW}=a\ket{0001}+b\ket{0010}+c\ket{0100}+d\ket{1000}$ where $a,b,c,d\in\mathbb{C}$ and $|a|^2+|b|^2+|c|^2+|d|^2=1$.\\
The negativities are 
$N_{1|234}=|d|\sqrt{|a|^2+|b|^2+|c|^2} $ , \\
$N_{2|134}=|c|\sqrt{|a|^2+|b|^2+|d|^2} $ , \\
$N_{3|124}=|b|\sqrt{|a|^2+|d|^2+|c|^2} $ , \\
$N_{4|123}=|a|\sqrt{|b|^2+|c|^2+|d|^2} $ , \\
$N_{12|34}=\sqrt{(|a|^2+|b|^2)(|c|^2+|d|^2)} $ , \\
$N_{13|24}=\sqrt{(|a|^2+|c|^2)(|b|^2+|d|^2)} $ , \\
$N_{14|23}=\sqrt{(|b|^2+|c|^2)(|a|^2+|d|^2)} $ . \\
$\delta_1=4|c|^2|d|^2,$ $\delta_2=4|a|^2|b|^2,$ $\delta_3=4|b|^2|d|^2,$ $\delta_4=4|a|^2|c|^2,$
$\delta_5=4|a|^2|d|^2,\delta_6=4|b|^2|c|^2$ and $\delta_7=0$. So $\delta_i\geq 0$ $\forall i=1,2,...,6$.\\
\\
Superposition of $\ket{GW}$ and $\ket{0000}$ is
$\ket{\psi}=\sqrt{p}\ket{GW}+\sqrt{1-p}\ket{0000}$ where $0<p<1$,$\ket{GW}=a\ket{0001}+b\ket{0010}+c\ket{0100}+d\ket{1000}$, $a,b,c,d\in\mathbb{C}$ s.t. $|a|^2+|b|^2+|c|^2+|d|^2=1 $. The Negativities are, 
$N_{1|234}=p|d|\sqrt{|a|^2+|b|^2+|c|^2} $ , \\
$N_{2|134}=p|c|\sqrt{|a|^2+|b|^2+|d|^2} $ , \\
$N_{3|124}=p|b|\sqrt{|a|^2+|d|^2+|c|^2} $ , \\
$N_{4|123}=p|a|\sqrt{|b|^2+|c|^2+|d|^2} $ , \\
$N_{12|34}=p\sqrt{(|a|^2+|b|^2)(|c|^2+|d|^2)} $ , \\
$N_{13|24}=p\sqrt{(|a|^2+|c|^2)(|b|^2+|d|^2)} $ , \\
$N_{14|23}=p\sqrt{(|b|^2+|c|^2)(|a|^2+|d|^2)} $ . \\
$\delta_1=4p^2|c|^2|d|^2,$ $\delta_2=4p^2|a|^2|b|^2,$ $\delta_3=4p^2|b|^2|d|^2,$ $\delta_4=4p^2|a|^2|c|^2,$
$\delta_5=4p^2|a|^2|d|^2,\delta_6=4p^2|b|^2|c|^2 $. So $\delta_i\geq 0$ $\forall i=1,2,...,6$.\\
\\
Superposition of  $\ket{GGHZ}$ and $\ket{W}$ state is\\
$\ket{\psi}=a\ket{0000}+b\ket{1111}+\frac{c}{2}(\ket{0001}+\ket{0010}+\ket{0100}+\ket{1000})$ where $a,b,c\in\mathbb{C}$ s.t. $|a|^2+|b|^2+|c|^2=1 \\ $.
$N_{1|234}=N_{2|134}=\sqrt{16|a|^2|b|^2+12|b|^2|c|^2+3|c|^4}/4=N_{3|124}=N_{4|123} \hspace{.1cm}, \\ 
N_{12|34}=\frac{|c|^2}{2}+\sqrt{2|a|^2|b|^2+2|b|^2|c|^2-2\sqrt{|a|^2|b|^4(|a|^2+2|c|^2)}}\\=N_{13|24}=N_{14|23} $ .\\
Since $N_{1|234}=N_{2|134}=N_{3|124}=N_{4|123} $ and $N_{12|34}=N_{13|24}=N_{14|23}$ we have, $\delta_i=N_{12|34}^2\geq0 \hspace{.1cm} \forall i=1,2,...,6$. 
\section*{Appendix 3}
\textbf{Theorem 1.} For an $N$ partite pure state $\ket{\psi_{A_1A_2...A_N}}$ in a $2\otimes2\otimes...\otimes2$($N$ times) system the negativity of bipartition $A_1|A_2...A_N$ is half of its concurrence, i.e., $N_{A_1|A_2...A_N}=\frac{1}{2}C_{A_1|A_2...A_N}$ \cite{10}. \\
\textbf{Proof.} For simplicity we write, $A_1=A$ and $A_2A_3...A_N=B$. By Schmidt decomposition, any bipartite state can be written as $\ket{\psi_{A|B}}=\sum_{i}\sqrt{\lambda_{i}}\ket{\phi_{A}^{i}}\otimes\ket{\phi_{B}^{i}}$ where $\lambda_i$ are Schmidt coefficients and $\{\ket{\phi_{A}^{i}}\},\{\ket{\phi_{B}^{i}}\}$ are orthogonal basis for the subsystems A and B.\\
Now, $\rho_{AB}=\sum_{i,j}\sqrt{\lambda_i\lambda_j}\ket{\phi_{A}^{i}}\bra{\phi_{A}^{j}}\otimes\ket{\phi_{B}^{i}}\bra{\phi_{B}^{j}}$\\
$\implies\rho_{AB}^{t_A}=\sum_{i,j}\sqrt{\lambda_i\lambda_j}\ket{\phi_{A}^{j'}}\bra{\phi_{A}^{i'}}\otimes\ket{\phi_{B}^{i}}\bra{\phi_{B}^{j}}$\\
So, we have\\
$N_{AB}=\frac{\|\rho_{AB}^{t_A}\|_1-1}{2} \\
= \frac{1}{2}\{\|\sum_{i,j}\sqrt{\lambda_i\lambda_j}\ket{\phi_{A}^{j'}}\bra{\phi_{A}^{i'}}\otimes\ket{\phi_{B}^{i}}\bra{\phi_{B}^{j}}\|_1-1\}\\
= \frac{1}{2}\{\|\sum_{i,j}\sqrt{\lambda_i\lambda_j}\ket{\phi_{A}^{j'}}\bra{\phi_{B}^{j}}\otimes\ket{\phi_{B}^{i}}\bra{\phi_{A}^{i'}}\|_1-1\}\\
= \frac{1}{2}\{\|\sum_j\sqrt{\lambda_j}\ket{\phi_{A}^{j'}}\bra{\phi_{B}^{j}}\otimes\sum_i\sqrt{\lambda_i}\ket{\phi_{B}^{i}}\bra{\phi_{A}^{i'}}\|_1-1\}\\
=\frac{1}{2}\{\|Z\otimes Z^{\dagger}\|_1-1\} \quad[$ $Z=\sum_{j=1}^{2}\sqrt{\lambda_j}\ket{\phi_{A}^{j'}}\bra{\phi_{B}^{j}}$ $ ] \\
= \frac{1}{2}\{\|Z\|^2_1-1\}\quad[$ $\|A\otimes B\|=\|A\|\|B\| $ $]\\
= \frac{1}{2}\{(\sqrt{\lambda_1}+\sqrt{\lambda_2})^2-1\}\\
= \frac{1}{2}\times2\sqrt{\lambda_1\lambda_2}\quad[$ $\sum_{i=1}^{2}\lambda_i=1 $ $]\\
= \frac{1}{2}\times2\sqrt{det(\rho_A)}\\
= \frac{1}{2}C_{AB}$\\
Hence, $N_{A_1|A_2...A_N}=\frac{1}{2}C_{A_1|A_2...A_N}$ (proved). \\

\textbf{Theorem 2.}
For an $N$ partite pure state $\ket{\psi_{A_1A_2...A_N}}$ in a $d_1\otimes d_2\otimes ...\otimes d_n$ dimensional system where $d_i>2$ $\forall i=1,2,...,n$, $N_{A_1|A_2...A_N}\geq\frac{1}{2}C_{A_1|A_2...A_N}$ .\\
\textbf{Proof.} For simplicity we write $A_1=A$ \& $A_2\otimes A_3\otimes...\otimes A_N=B$. Suppose, $d\leq min\{d_1, d_2.d_3...d_n\}$, then by Schmidt decomposition for any bipartite state, we write, $\ket{\Psi_{A|B}}=\sum_{i=1}^{d}\sqrt{\lambda_{i}}\ket{\phi_{A}^{i}}\otimes\ket{\phi_{B}^{i}}$ where $\lambda_i$ are Schmidt coefficients and $\{\ket{\phi_{A}^{i}}\},\{\ket{\phi_{B}^{i}}\}$ are orthogonal basis for the subsystems A and B respectively. By the similar calculations from theorem 1 we can say that\\
$N_{AB} = \frac{1}{2}\{\|Z\|^2_1-1\}
= \frac{1}{2}\{[\sum_{i=1}^{d}\sqrt{\lambda_i}\hspace{3pt} ]^2-1\}\\
= \frac{1}{2}(2\sum_{i\neq j=1}^{\binom{d}{2}}\sqrt{\lambda_i\lambda_j})\\
\geq \frac{1}{2}\times2\times \binom{d}{2}\sqrt{\prod_{i=1}^{d}\lambda_i}\geq \frac{1}{2}\times2\sqrt{\prod_{i=1}^{d}\lambda_i}\\
\implies N_{AB} \geq \frac{1}{2}\times2\sqrt{\lambda_1\lambda_2...\lambda_{d}}\\
\implies N_{AB} \geq \frac{1}{2}\times2\sqrt{det(\rho_A)} \\
\implies N_{AB} \geq \frac{1}{2}C_{AB}$ \\
where $Z=\sum_{i=1}^{d}\sqrt{\lambda_i}\ket{\phi_{A}^{i}}\bra{\phi_{B}^{i}},\hspace{2pt} \|A\otimes B\|=\|A\|\|B\|$ \hspace{2pt} and $\sum_{i=1}^{d}\lambda_i=1$\\
Hence, $N_{A_1|A_2...A_N}\geq\frac{1}{2}C_{A_1|A_2...A_N}$ (proved).\\


\begin{thebibliography}{4}
	\bibitem{2}Bennett, C.H., Wiesner, S.J. : Communication via one- and two-particle operators on Einstein-Podolsky-Rosen states. Phys. Rev. Lett. \textbf{69}, 2881 (1992) 
	\bibitem{3} Bennett, C.H., Brassard, G., Crepeau, C., Jozsa, R., Peres, A., Wooters, W.K.: Teleporting an unknown quantum state via dual classical and Einstein-Podolsky-Rosen channels. Phys. Rev. Lett. \textbf{70}, 1895 (1993)
	\bibitem{4} Bennett, C.H., Divincenzo, D.P.: Quantum information and computation. Nature(London) \textbf{404}, 247 (2000)
	\bibitem{5} Ruussendorf, R., Briegel, H.J.: A One-Way Quantum Computer. Phys. Rev. Lett. \textbf{86}, 5188 (2001)
	\bibitem{sev} Seevinck, M.P.: Monogamy of correlations versus monogamy of entanglement. Quantum Inf. Process, \textbf{9}, 273-294 (2010)
	\bibitem{6}  Coffman, V., Kundu, J., Wootters, W.K.: Distributed entanglement. Phys. Rev. A  \textbf{61}, 052306 (2000)
	\bibitem{7} Osborne, T.J., Verstraete, F.: General Monogamy Inequality for Bipartite Qubit Entanglement. Phys. Rev. Lett. \textbf{96}, 220503 (2006)
	\bibitem{8} Regula, B., Martino, S.D., Lee, S., Adesso, G.: Strong Monogamy Conjecture for Multiqubit Entanglement: The Four-Qubit Case. Phys. Rev. Lett. \textbf{113}, 110501 (2014)
	\bibitem{9} Karmakar, S., Sen, A., Bhar, A., Sarkar, D.: Strong monogamy conjecture in a four-qubit system. Phys. Rev. A \textbf{93}, 012327 (2016)
	\bibitem{10} Luo, Y., Li, Y.: Monogamy of $\alpha$-th power entanglement measurement in qubit systems. Ann. Phys. \textbf{362}, 511-520 (2015)
	\bibitem{11} He, H., Vidal, G.: Disentangling theorem and monogamy for entanglement negativity. Phys. Rev. A \textbf{91}, 012339  (2015)
	\bibitem{12} Ou, Y.C., Fan, H.: Monogamy inequality in terms of negativity for three-qubit states. Phys. Rev. A \textbf{75}, 062308 (2007)
	\bibitem{13} Lancien, C., Martino, S.D., Huber, M., Piani, M., Adesso, G., Winter, A.: Should Entanglement Measures be Monogamous or Faithful?. Phys. Rev. Lett. \textbf{117}, 060501 (2016)
	\bibitem{14} Gour, G., Guo, Y.: Monogamy of entanglement without inequalities. Quantum \textbf{2}, 81 (2018)
	\bibitem{1} Eltschka, C., Huber, F., G\"uhne, O., Siewert, J.: Exponentially many entanglement and correlation constraints for multipartite quantum states. Phys. Rev. A \textbf{98}, 052317 (2018)
	\bibitem{15}  Vide, G., Werner R.F.: Computable measure of entanglement. Phys. Rev. A \textbf{65}, 032314 (2002)
	\bibitem{16} Peres, A.: Separability Criterion for Density Matrices. Phys. Rev. Lett. \textbf{77}, 1413 (1996)
	\bibitem{motivation 1} Rains, E.M.: Quantum weight enumerators. IEEE Trans. Inf. Theory 44, 1388 (1998)
	\bibitem{motivation 2} Rains, E.M.: Polynomial invariants of quantum codes. IEEE Trans. Inf. Theory \textbf{46}, 54 (2000)
	\bibitem{motivation 3}  Huber, F., Eltschka, C., Siewert, J., G\"uhne O.: Bounds on absolutely maximally entangled states from shadow inequalities, and the quantum MacWilliams identity. J. Phys. A \textbf{51}, 175301 (2018)
	\bibitem{17} Verstraete, F., Dehaene, J., Moor, B.D., Verschelde, H.: Four qubits can be entangled in nine different ways. Phys. Rev. A \textbf{65}, 052112 (2002)
	\bibitem{18}  Gour, G., Wallach, N.R.: All maximally entangled four-qubit states. J.Math.Phys. \textbf{51}, 112201 (2010)
	\bibitem{19}  G\"uhne, O., T\'oth, G., Hyllus, P., Briegel, H.J.: Bell Inequalities for Graph States. Phys. Rev. Lett. \textbf{95}, 120405 (2005)
	\bibitem{20}  Schlingemann, D., Werner, R.F.: Quantum error-correcting codes associated with graphs. Phys. Rev. A \textbf{65}, 012308 (2001)
	\bibitem{21}  Bai, Y.K., Wang, Z.D.: Multipartite entanglement in four-qubit cluster-class states. Phys. Rev. A \textbf{77}, 032313 (2008)
	\bibitem{22}  Dicke, R.H.: Coherence in Spontaneous Radiation Processes. Phys. Rev. \textbf{93}, 99 (1954)
\end{thebibliography}
\end{document}